\documentstyle[preprint,aps,prc]{revtex}
\draft
\begin{document}

\title{Nucleon Helicity in Pion Photoproduction}
\author{D. A. Jenkins and I. I. Strakovsky}
\address{Physics Department,\\
 Virginia Polytechnic Institute and State University,\\
 Blacksburg, Virginia, USA. 24061-0435\\}
\date{\today}
\maketitle
\begin{abstract}
Pion-photoproduction data is examined to check for the nucleon-helicity
conservation predicted by asymptotic QCD.  The differential cross
section shows agreement with constituent-counting rules, and
polarization
data is not in disagreement with conservation of nucleon helicity.
However large uncertainties in the polarization measurements do not
allow a conclusive statement.  The helicity amplitudes from a
partial-wave
analysis are also examined for helicity conservation.   While the
amplitudes become small as $s$ increases, the $s$ dependence of
the helicity-conserving amplitudes is similar to the dependence of the
non-conserving amplitudes.
\end{abstract}

\newpage

The application of QCD to the study of nuclear phenomena is an important
issue which will be studied by the new generation of several-GeV,
high-intensity electron accelerators.  The present data for pion
photoproduction and deuteron photodisintegration offer a means for a
first look at these issues.

The question of the applicability of perturbative QCD to deuteron
photodisintegration has been addressed recently by Belz {\it et al.}
who measured
the differential cross section for energies up to 2.8 GeV.\cite{Be95}
Their results
demonstrate that at $\theta_{cm}$ = 90$^\circ$
the cross section is in good agreement with
constituent-counting-rule predictions for incident-photon energies greater
than 1.5 GeV.  These results are surprising in view of the observation of
Isgur and Llewellyn-Smith that perturbative contributions to the cross
section are small compared to other processes.\cite{Is89}
Experiments at higher energies
may provide further insight into this issue.\cite{Ho89}
Another prediction of asymptotic
QCD, the conservation of nucleon helicity\cite{Br89},
will be tested by measurements of
polarization observables.\cite{Gi94}

Compared to deuteron photodisintegration, there is a much larger set of data
for pion photoproduction which can be examined for evidence of nucleon-helicity
conservation.  The current data from the SAID data base
for the differential cross section of the
$\gamma p \rightarrow \pi^+n$ reaction channel is shown in
Figure 1a.\cite{SAID}
As has been observed by
Anderson {\it et al.}, the reaction shows
agreement with constituent counting rules that
predict the cross section should vary as $s^{-7}$.\cite{An76}
The agreement extends down to
700 MeV photon energy ($s$ = 2.2 GeV$^2$) where baryon
resonances are important. Oscillations about the counting-rule prediction
are similar to those observed in elastic pp scattering at
90$^\circ$.\cite{Ak67}

There are three polarization observables for pion photoproduction
which vanish if nucleon helicity is
conserved.
These are the recoil polarization asymmetry $P$, the target-polarization
asymmetry $T$ and the double-polarization observable for polarized photons on
a transversely-polarized target $H$.
Plots of these observables at 90$^\circ$ are
shown in Fig. 1 for $P$ (Fig. 1b) in the
$\gamma p \rightarrow \pi^\circ p$ channel and $T$ (Fig. 1c) and $H$ (Fig. 1d)
in the
$\gamma p \rightarrow \pi^+ n$
channel.  The observables become
small as $s$ increases, but a definite conclusion is difficult to draw because
of large uncertainties in the data. The polarization
data extend up to $s \simeq$
4 GeV$^2$.
Data in
other pion channels is less extensive, but do not show any different trends.

A more direct means of checking helicity conservation is to look at the
helicity amplitudes themselves, as obtained from a partial-wave analysis.
A recent partial-wave analysis fits
the data up to a photon energy of 2 GeV.\cite{Ar95}
The partial-wave results for the helicity amplitudes reflect all of
the measured observables for the reaction and should, in principle,
be better determined
than a polarization observable which can have large uncertainties.

Pion photoproduction
is described by eight helicity amplitudes, but with parity conservation only
four of the amplitudes are independent.  The partial-wave analysis
is performed by expanding each of the helicity amplitudes in terms of multipole
amplitudes which depend on energy.  The amplitudes are varied in a $\chi^2$
fitting
procedure to give a best fit to all measured
observables.  Since the photoproduction
data is not sufficiently complete for a fit to the data, the fit is
constrained by relations such as Watson's theorem and dispersion
relations.
Consequently the results depend on the constraining relations and are
model
dependent.  The situation will be improved by new data from
CEBAF.\cite{Br94}

The four nucleon helicities for the $\gamma p\rightarrow \pi^\circ p$
reaction channel as determined from partial-wave
analysis are shown in Figure 2.  Two of the helicity amplitudes, $H_2$ and
$H_3$ in
the notation of Walker\cite{Wa69},
should be zero if nucleon helicity is conserved.
As $s$ increases, all four helicity amplitudes become small.  The helicity
conserving amplitudes
are somewhat
larger than the
amplitudes which do not conserve helicity.

In conclusion there is no strong evidence for helicity conservation in the
pion-photoproduction data although the polarization data suggests the approach
of helicity conservation as energy increases.  More data is needed at higher
energies.
Improved data is also needed in order to reduce the
constraints for the partial-wave analysis and extend the analysis to higher
energies. The tagged photon beam at CEBAF will provide data with small angular
intervals over a large range of angles and energies and will be free of
normalization uncertainties associated with much of the present data which
has been taken with bremsstrahlung beams.

We thank R. A. Arndt and B. Z. Kopeliovich for helpful discussions.

This work was supported in part by the National Science Foundation under grants
NSF-PHY-9207000
and the
U. S. Department of Energy under grant DE-FG05-88ER40454.

\newpage

\begin{figure}
\caption{Differential cross section and polarization observables for pion
photoproduction at 90$^\circ$ c. m. angle.
The
data are taken from the SAID data base.\protect\cite{SAID}.
(a) The
differential cross section $d\sigma /dt$ versus $s$ for the
reaction $\gamma p \rightarrow \pi^+ n$.
The dashed line shows the function $s^{-7}$ for
reference.
(b) P, the recoil-polarization
asymmetry for the $\gamma p \rightarrow \pi^\circ p$
reaction, c) T, the target-polarization
asymmetry
for the $\gamma p \rightarrow \pi^+ n$ reaction, and d) H, the
double-polarization observable for polarized photons on a
transversely-polarized
target for the $\gamma p \rightarrow \pi^+ n$ reaction.
The polarization observables are equal to zero if the reaction
conserves nucleon helicity.
The solid
line shows the prediction SM95 from the partial-wave analysis of Arndt
et al.\protect\cite{Ar95}}
\end{figure}

\begin{figure}
\caption{Helicity amplitudes for pion photoproduction
versus $s$ for the $\gamma p \rightarrow \pi^\circ p$ reaction
at 90$^\circ$ c.m. angle as determined in the SM95 solution of the
analysis of Arndt
{\it et al.}\protect\cite{Ar95}.
(a) $H_1$, (b) $H_2$,
(c) $H_3$ and (d)$H_4$.  Solid (dashed) curves give the real (imaginary)
parts of the amplitudes.
If the reaction conserves helicity,
$H_2$ and $H_3$ should equal zero.}
\end{figure}


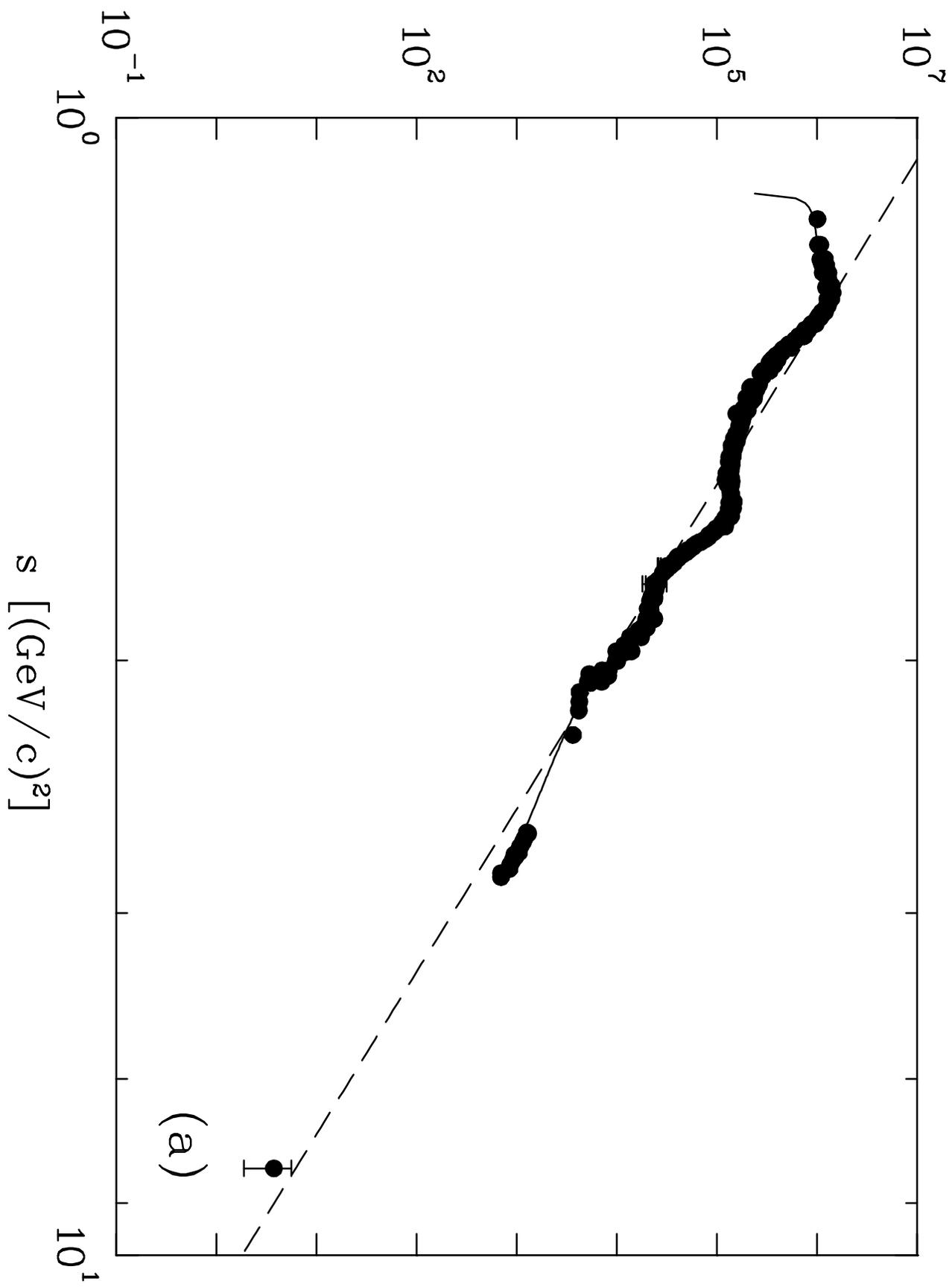
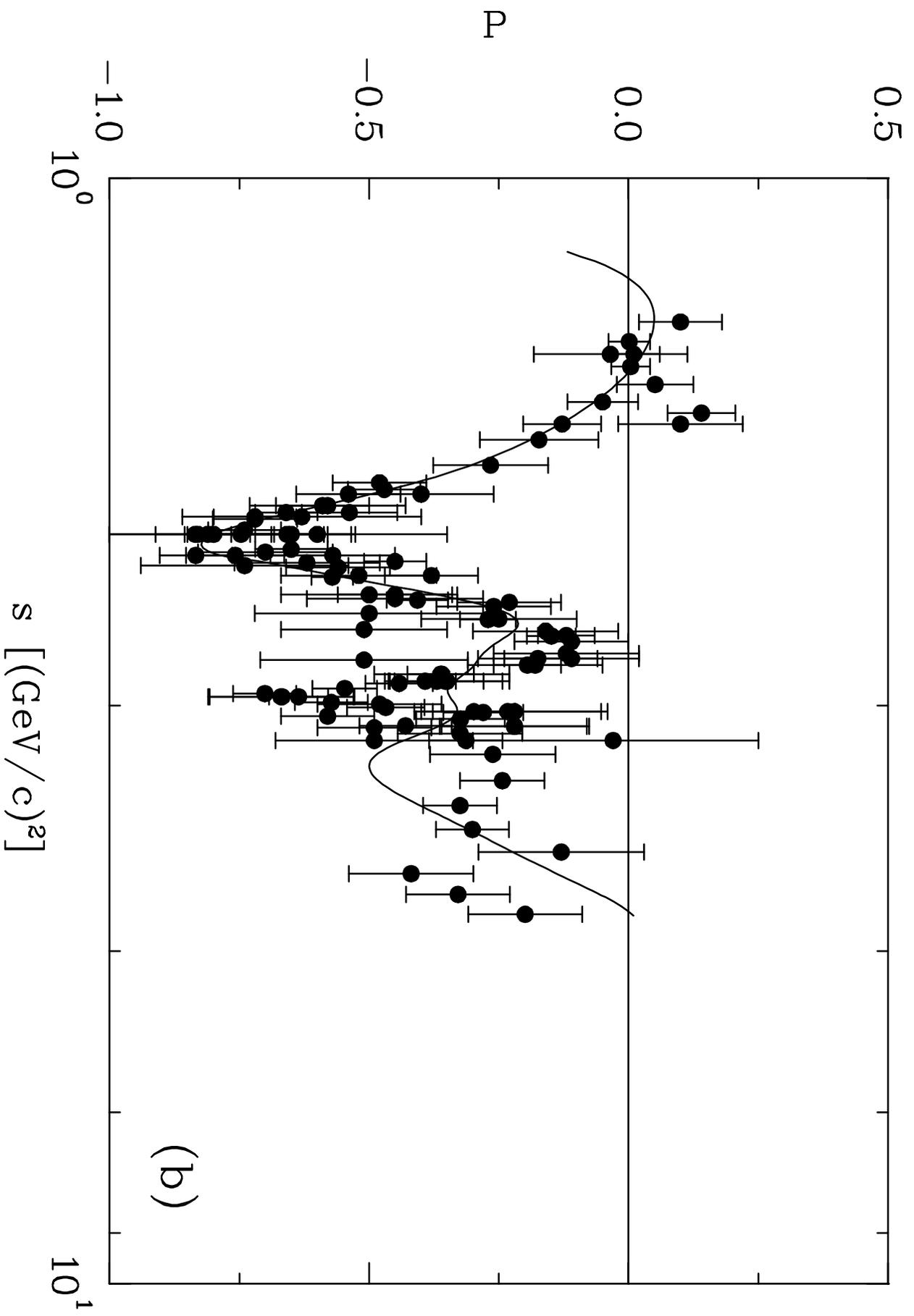
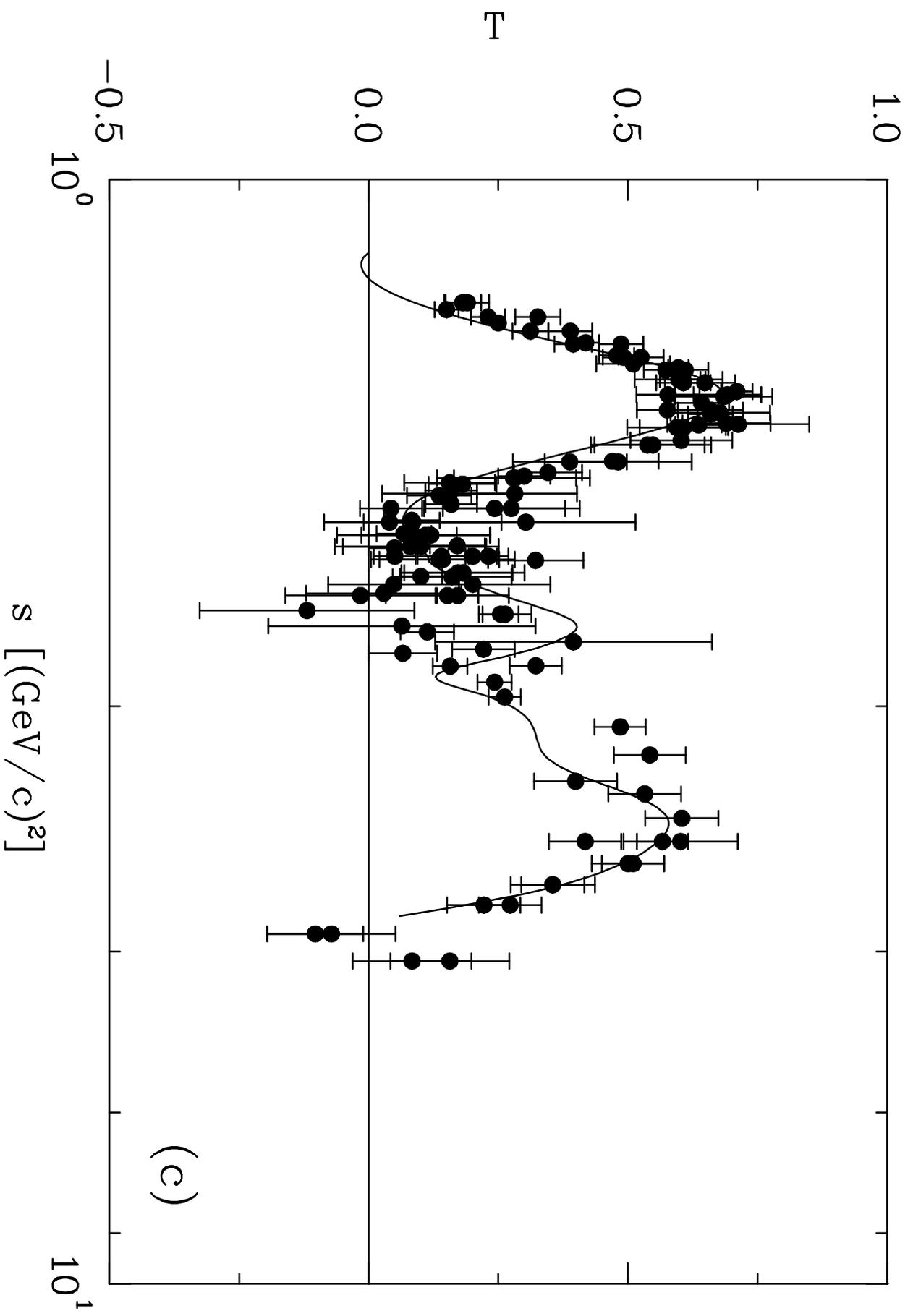
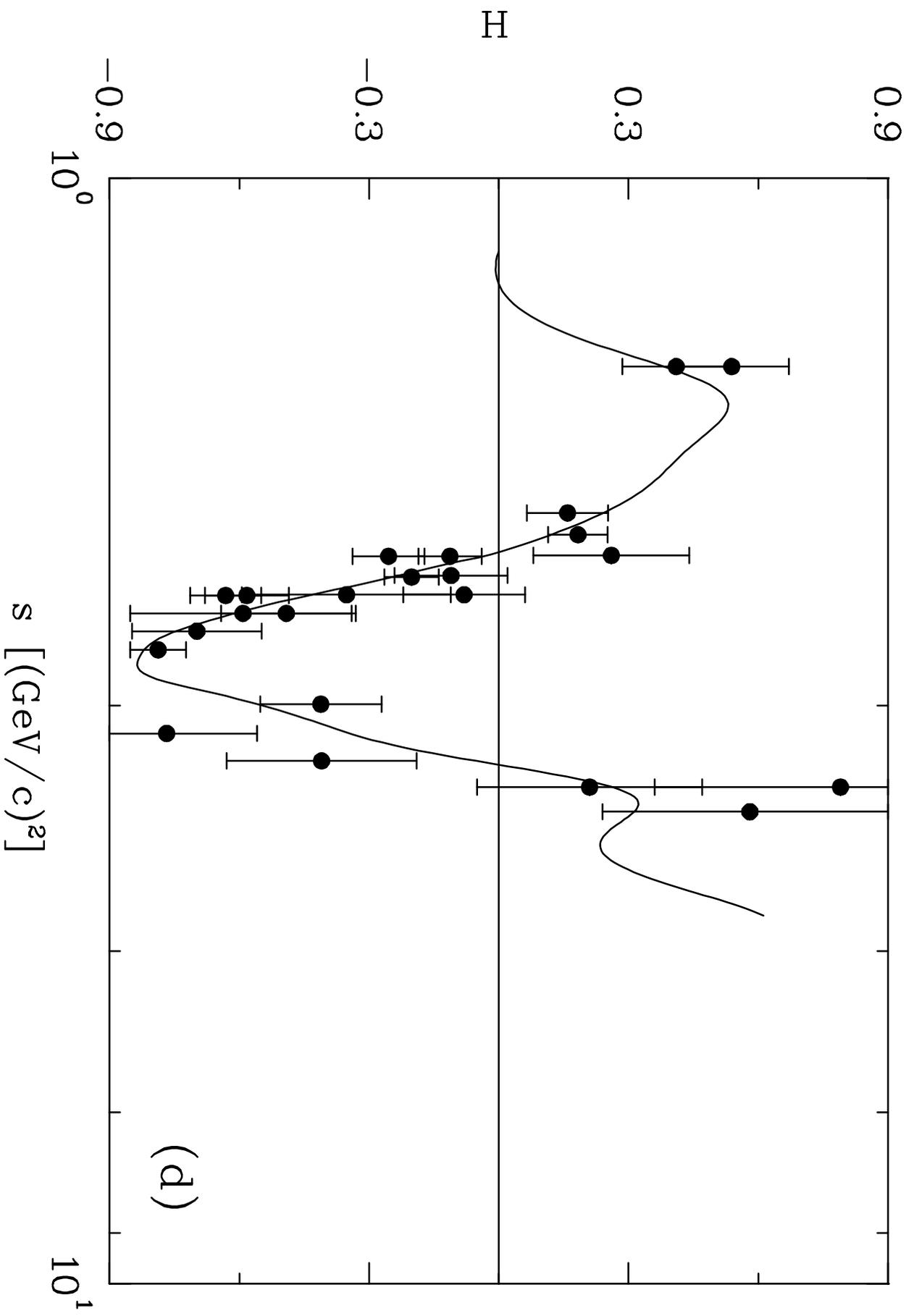
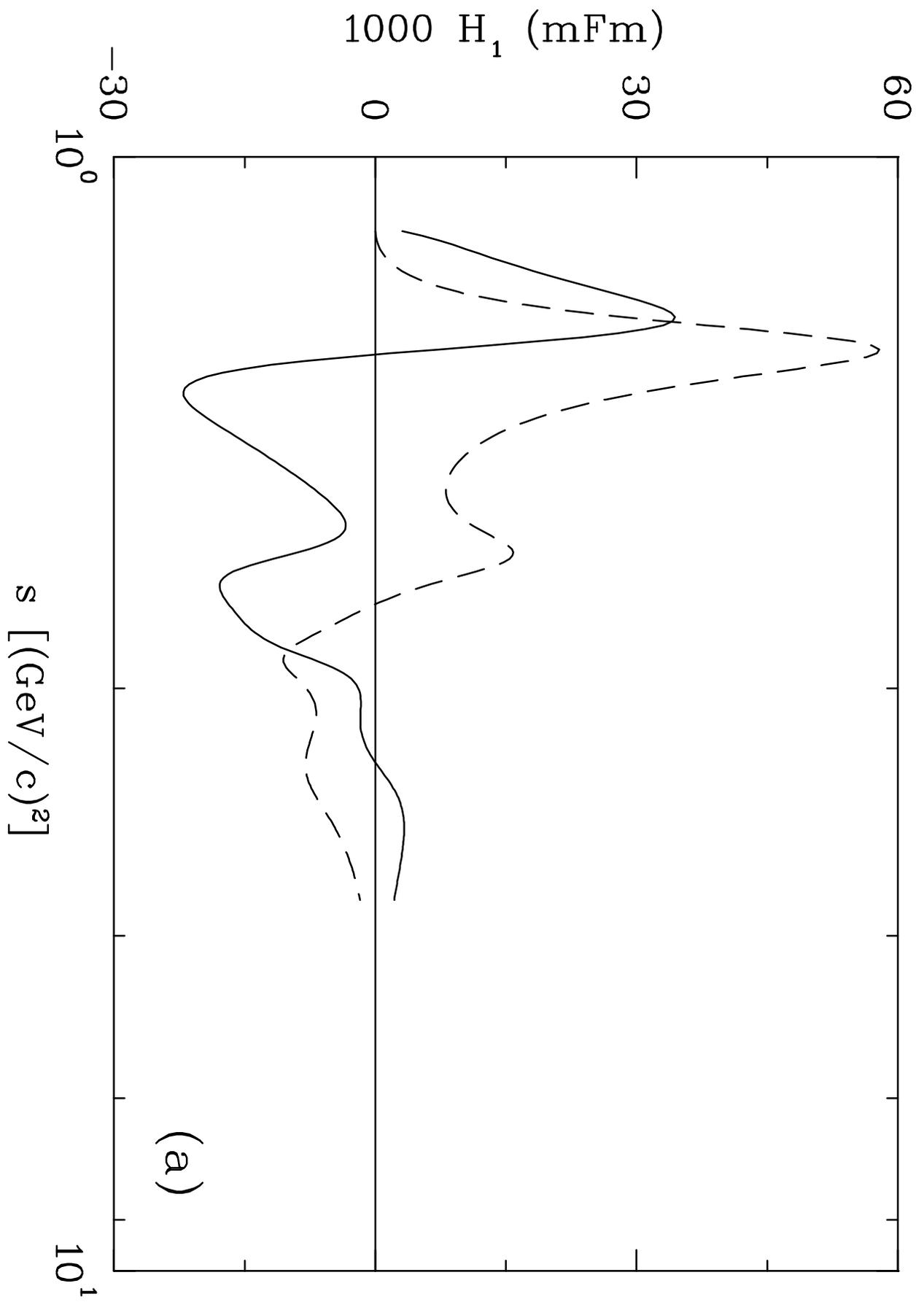
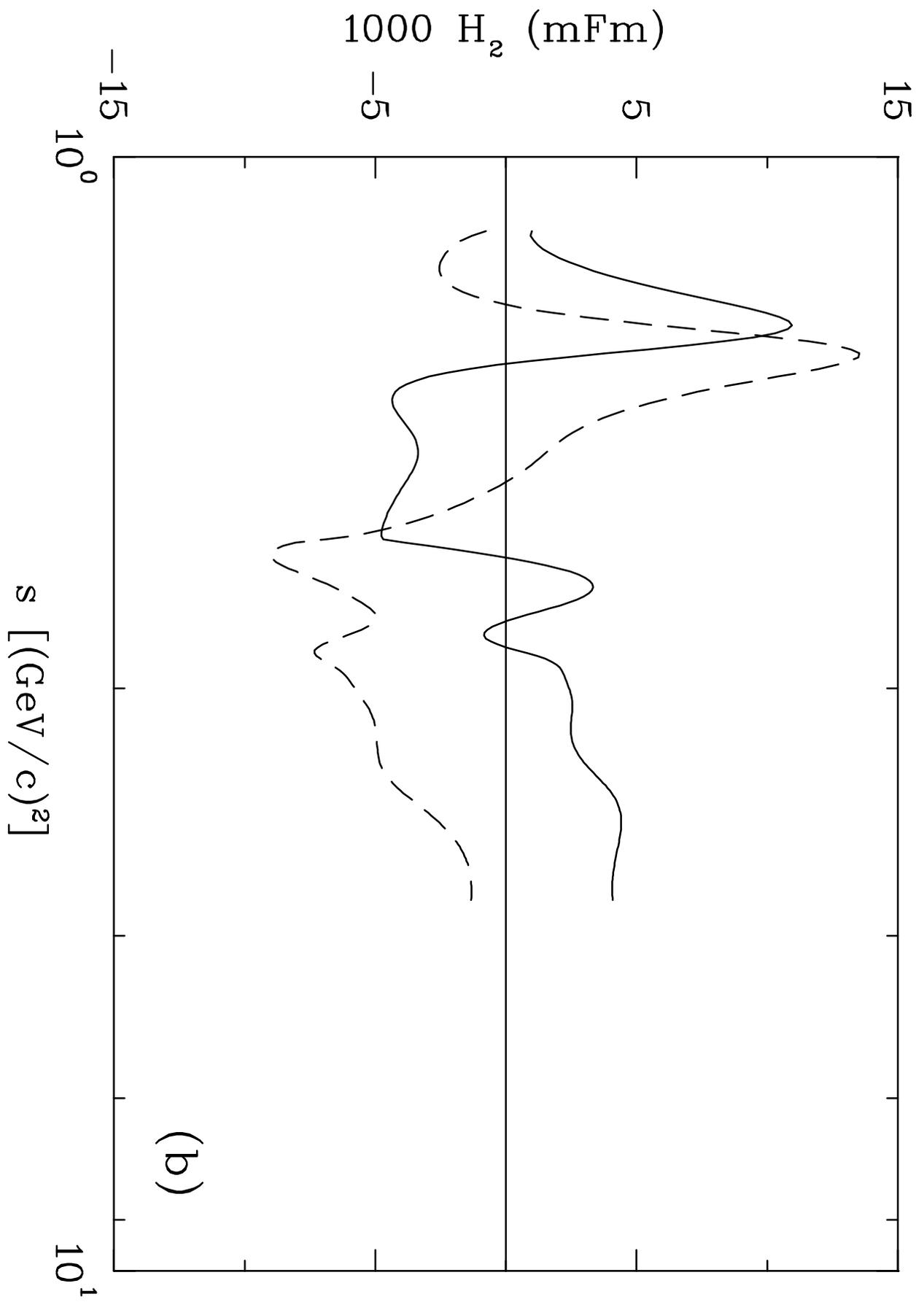
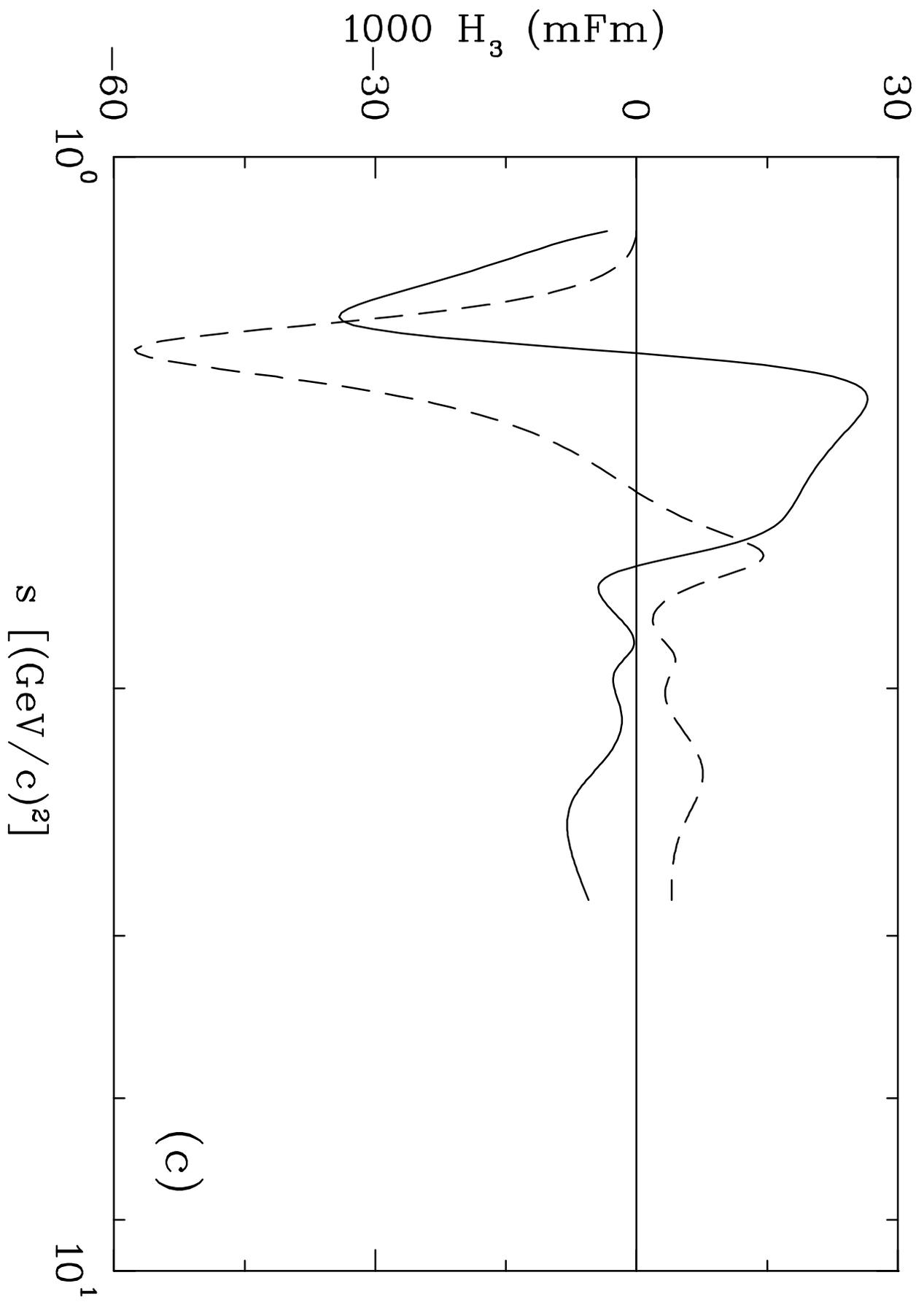
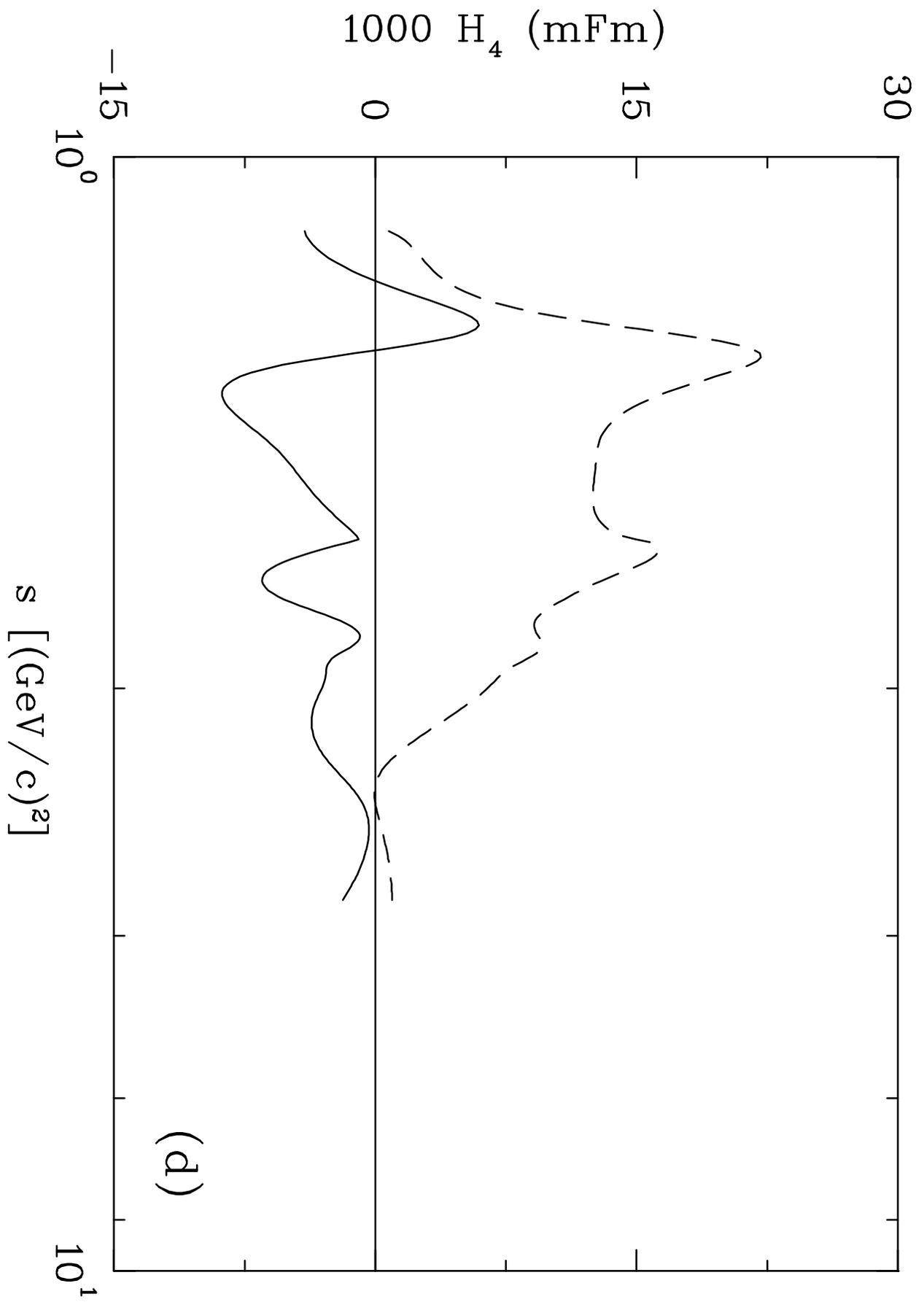
\end{document}